\documentclass{PoS}

\usepackage{overpic}

\newcommand{\phiphi}{\ensuremath{B_s^0 \to \varphi \varphi}}
\newcommand{\phikk}{\ensuremath{\varphi \to K^+K^-}}
\newcommand{\jpsi}{\ensuremath{ J\!/\!\psi}}
\newcommand{\fnot}{\ensuremath{f_0(980)}}
\newcommand{\fnotpipi}{\ensuremath{f_0(980)\to \pi^+\pi^-}}

\newcommand{\jpsiphi}{\ensuremath{B_s^0 \to J\!/\!\psi \varphi}}
\newcommand{\jpsifnot}{\ensuremath{B_s^0 \to J\!/\!\psi f_0(980)}}
\newcommand{\bsjpsik}{\ensuremath{B_s^0 \to J\!/\!\psi K^{(\star)}}}
\newcommand{\bsjpsikst}{\ensuremath{B_s^0 \to J\!/\!\psi K^{\star}(892)^0}}
\newcommand{\bsjpsikS}{\ensuremath{B_s^0 \to J\!/\!\psi K_S}}
\newcommand{\bjpsix}{\ensuremath{B \to J\!/\!\psi X}}
\newcommand{\phikst}{\ensuremath{B^0 \to \varphi K^{\star}(892)^0}}

\newcommand{\betas}{\ensuremath{\beta_s}}
\newcommand{\CP}{\ensuremath{\mathsf{CP}}}
\newcommand{\DGs}{\ensuremath{\Delta \Gamma_s}}
\newcommand{\Br}{\ensuremath{\mathcal{B}}}

\newcommand{\bs}{\ensuremath{B_s^0}}

\newcommand{\pt}{\ensuremath{p_{\rm{T}}}}
\newcommand{\ifb}{fb$^{-1}$}

\newcommand{\au}{\ensuremath{\mathcal{A}_u}}
\newcommand{\av}{\ensuremath{\mathcal{A}_v}}

\newcommand{\stat}{\ensuremath{\rm{(stat.)}}}
\newcommand{\syst}{\ensuremath{\rm{(syst.)}}}

\title{Suppressed \bs\ decays at CDF}

\ShortTitle{Suppressed Bs decays at CDF}

\author{Mirco Dorigo
       \thanks{Speaker on behalf of the CDF collaboration.}\\
       INFN and University of  Trieste\\
       E-mail: \email{mirco.dorigo@ts.infn.it}}

\abstract{We review three recent results of the CDF
  collaboration on \bs\ suppressed decays:  the first search for \CP--violation in the
  \phiphi\ decay, where two \CP--violating asymmetries expected to be zero in the
  Standard Model are measured, and the observation and the branching
  ratio measurements of \jpsifnot\ and \bsjpsik\ decays.
}

\FullConference{The 13th International Conference on B-Physics at Hadron Machines\\
		 April 4-8 2011\\
		 Amsterdam, The Netherlands}

\begin{document}

\section{Introduction}
In the past decade Tevatron experiments CDF and
D0 have pioneered  the physics of the \bs\ meson with
a broad program aimed at its exploration.
Although significant samples of fully reconstructed \bs\ decays have been collected
allowing decisive progress on \bs\ mixing, lifetime, decay width
difference \DGs\ as well as \CP--violation measurements,
 more precise investigations are ongoing. 
In this report, we review recent results of the CDF collaboration
on \bs\ suppressed decays: the first search for \CP--violation in the
\phiphi\ decay and the observation of \jpsifnot\ and \bsjpsik\ decays.

Two features of the CDF II detector~\cite{CDFdetector} are relevant
for these measurements: the tracking and the trigger. 
A high resolution tracking
detector provides an excellent resolution on $B$--meson decay length (30 $\mu$m) and mass, 
typically about 10 MeV/$c^2$ for \bjpsix\ modes, that is pivotal for the
observation of  \bs\ suppressed modes. This is achieved by double-sided silicon microstrips arranged in five
cylindrical layers 
and an open cell drift chamber with 96 sense
wires, 
all immersed in a 1.4 T solenoidal magnetic field.
Signals of \bjpsix\ modes are efficiently collected by a dimuon
trigger~\cite{CDFdetector} with a 1.5 GeV/$c$ transverse momentum
threshold, 
 while the trigger on displaced vertex~\cite{SVT}
 allows the collection of hadronic decay modes like \phiphi, 
through online measurement of impact parameters of charged tracks with a resolution
(48 $\mu$m) comparable with offline measurements.

\section{First search for \CP--violation in the \phiphi\ decay}
The $\phiphi$  decay belongs to the class of transitions of pseudoscalar mesons 
into two vector particles ($P\to VV$), whose rich dynamics involves 
three different amplitudes corresponding to the
polarization states.  In the Standard Model (SM) the dominant quark level process is
described by the $b\rightarrow s$
``penguin'' amplitude. Hence, the possibility to access New Physics (NP)
through exchange of new virtual massive particles makes the \phiphi\ channel attractive.
Indeed,  the na\"ive SM expectation for polarization
amplitudes has shown discrepancies with measurements of
 similar penguin decays~\cite{BVV_exp}, raising considerable attention to the so--called
``polarization puzzle''~\cite{BVV_th}.
Moreover, having a self--conjugate final state, the \phiphi\ 
decay is sensitive to the \CP--violation in the
interference between decay with and without mixing. 
 Actually, the \CP--violating weak phase $\phi_s^{\phiphi}$ is predicted to be
 extremely small in the SM  
and measurement of nonzero \CP--violating observables  
 would indicate unambiguously NP.

The first evidence for the $\phiphi$ decay has been reported by CDF 
in 2005~\cite{phiphi_PRL}.  Using 2.9 fb$^{-1}$ of data, the branching ratio
measurement was recently updated~\cite{phiphi},
$\Br(\phiphi)=(2.40 \pm 0.21\stat \pm 0.86\syst)
\times 10^{-5}$,
in agreement with the first determination.
Signal candidates are reconstructed by detecting \phikk\ 
decays and are formed by fitting four tracks to a common vertex. 
Combinatorial background is reduced by exploiting several
variables sensitive to the long lifetime and relatively
hard \pt\ spectrum of $B$ mesons, while the physics background, given by \phikst\ decay, is estimated by simulation not to exceed a
3\% fraction of the signal. Signals of $295\pm20$ events are obtained by fitting the mass distribution.
This data sample has allowed the world's first 
polarization measurement~\cite{phiphi} by analyzing the  angular distributions of decay products, expressed as a function of helicity angles,
$\vec{\omega}=(\cos\vartheta_1,\cos\vartheta_2,\Phi$). 
The total decay width is composed of three polarization amplitudes:
two \CP--even ($A_0$ and $A_\parallel$) and one \CP--odd ($A_\perp$). The measured amplitudes result in a
smaller longitudinal fraction
with respect to the na\"ive expectation, $f_{\rm{L}}=0.348\pm0.041\pm0.021$, 
 as found in other similar $b \to s$ penguin
decays~\cite{BVV_exp}.

\begin{figure} 
\begin{center}
\begin{overpic}[width=0.35 \columnwidth]{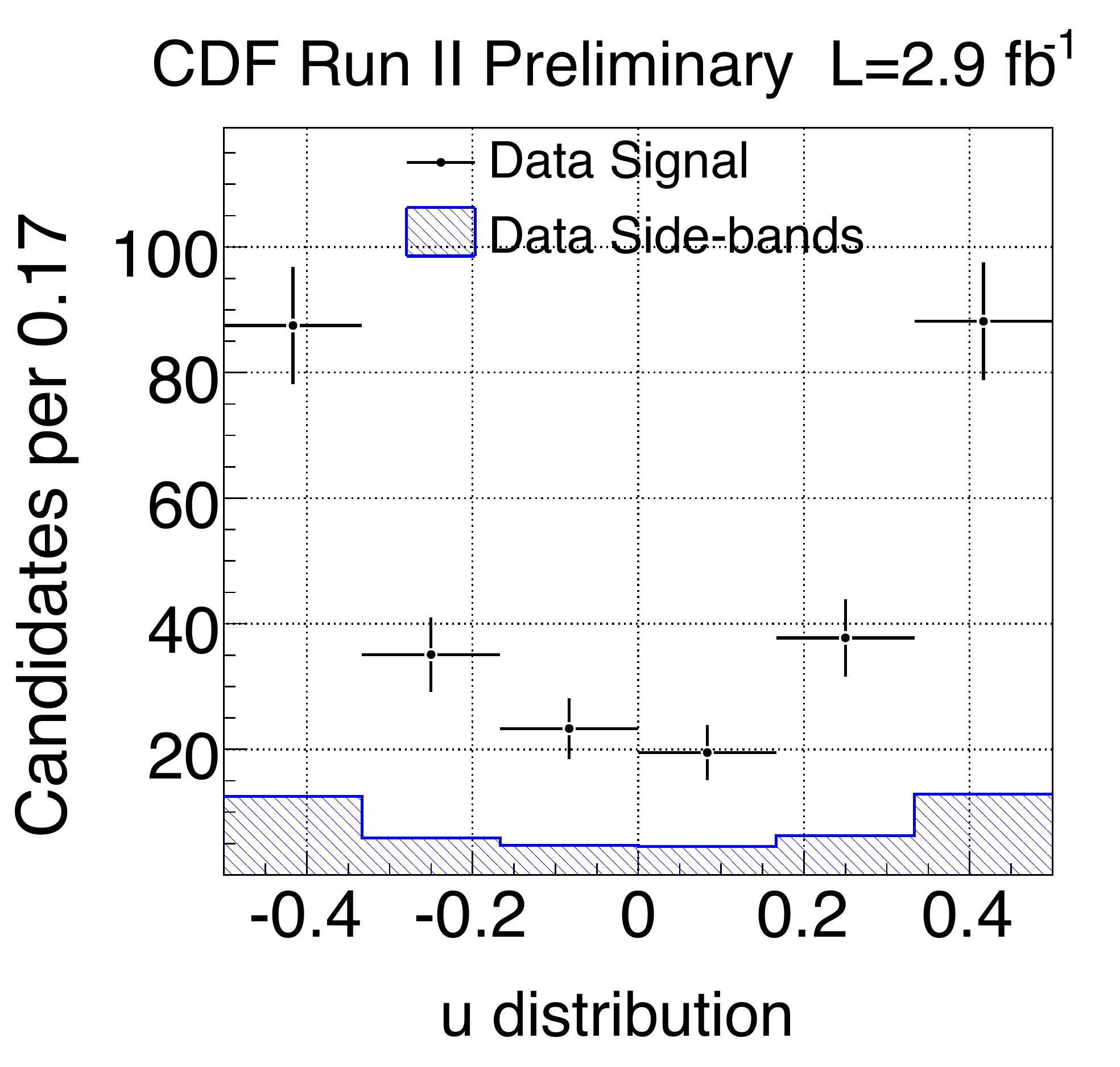}
\end{overpic}
\begin{overpic}[width=0.35\columnwidth]{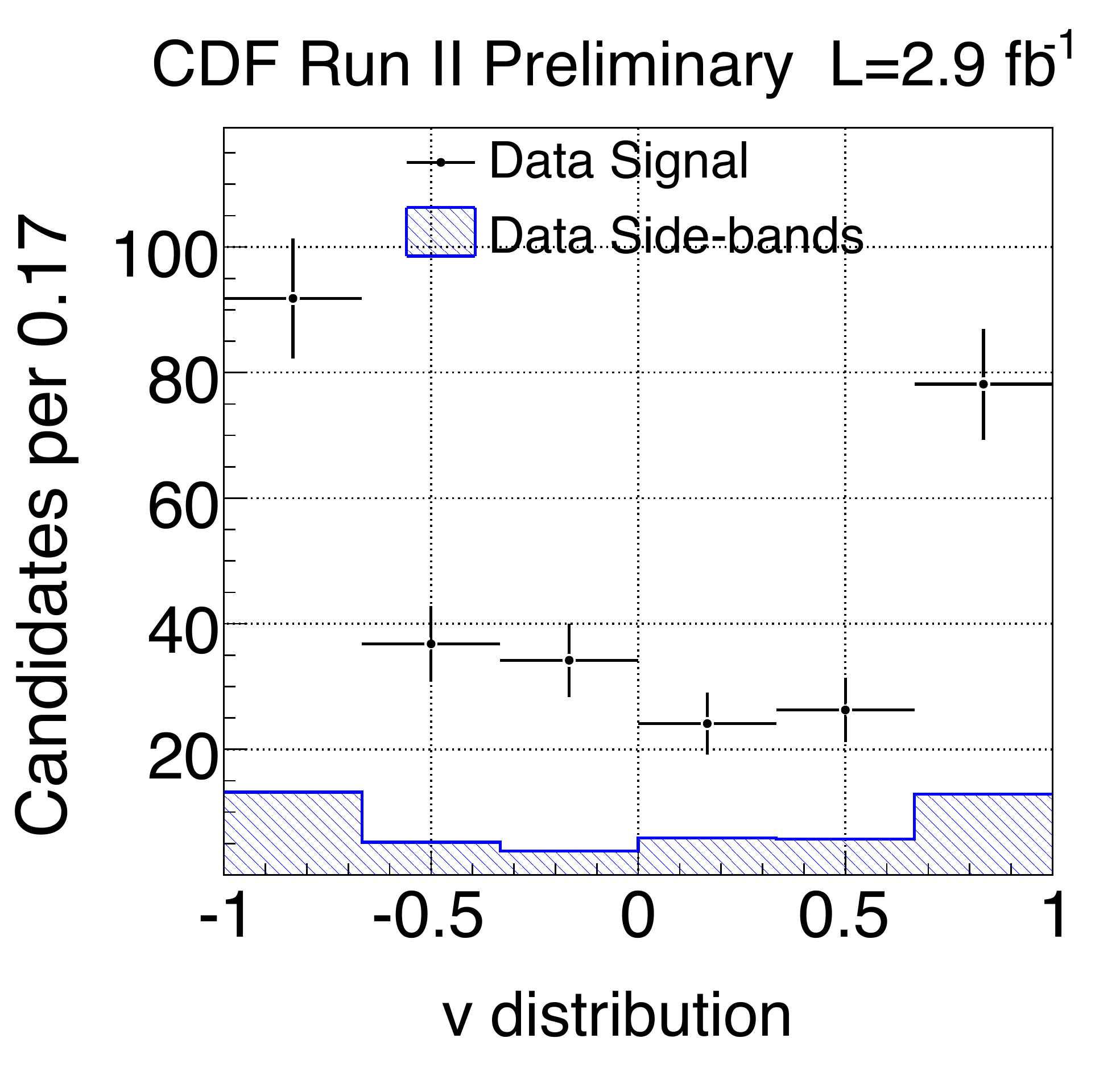}
\end{overpic}
\end{center}
\caption{\label{fig:TP}
Distribution of $u$ (left) and $v$ (right) for \phiphi\ candidates. Black crosses
are background--subtracted data; the blue histogram represents the background.}
\end{figure}

Present statistics of the \phiphi\ data sample are not sufficient for 
a suitable time--dependent analysis of mixing induced \CP--violation as
the case of the \jpsiphi\ decay. However, an investigation of
genuine \CP--violating observables which could reveal the presence of NP, such as triple
products (TP) correlations, is accessible~\cite{TP_th}.  The TP is
expressed as $\vec{p} \cdot (\vec{\epsilon}_1 \times \vec{\epsilon}_2)$, where
$\vec{p}$ is the momentum of one of the $\varphi$ meson in the \bs\ rest frame,
and $\vec{\epsilon}_i$ are the polarization vectors of the vector mesons. 
There are two triple products in the \phiphi\ decay corresponding to 
interferences between \CP--odd and \CP--even amplitudes, one for
transverse--longitudinal mixture, $\Im(A_0A_\perp^\star)$, and the other for the
transverse--transverse term, $\Im(A_\parallel A_\perp^\star)$. These
 products are functions of the helicity angles: the former is defined by
$v=\sin\Phi$ for $\cos\vartheta_1 \cos\vartheta_2\geq 0$ and $v=-\sin\Phi$
for $\cos\vartheta_1 \cos\vartheta_2<0$; the latter is defined by
$u=\sin2\Phi$. The $u$ and $v$ distribution for \phiphi\ candidates are shown in fig~\ref{fig:TP}.
 Without distinction of the flavor of the \bs\ meson at the production
time (\emph{untagged} sample), the following equation defines a \CP--violating asymmetry:
\begin{equation}
\au=\frac{\Gamma(u>0) + \bar{\Gamma}(u>0) -\Gamma(u<0) - \bar{\Gamma}(u<0)}
{\Gamma(u>0) + \bar{\Gamma}(u>0) +\Gamma(u<0) + \bar{\Gamma}(u<0)},
\end{equation}
where $\Gamma$ is the decay rate for the given process and
$\bar{\Gamma}$ is its \CP--conjugate. An equivalent definition holds
for $v$. Being proportional to $\sin\phi_s\cos\delta_i$, where $\delta_i$ are
relative strong phases between the polarization amplitudes, in
\phiphi\ these asymmetries are nonzero only in presence of NP~\cite{TP_th}.

The CDF collaboration has made the first measurement of \au\ and \av\
asymmetries in \phiphi\ using the data sample described above~\cite{TP_CDF}.
The asymmetries are obtained
through an unbinned maximum likelihood fit. 
The sample is split into two subsets
according to the sign of $u$ (or $v$) of \phiphi\ candidates. 
The invariant mass distribution of each subset is fitted
simultaneously in order to extract the signal asymmetry.  
The small fraction of physics background, such as \phikst\ as well as non--resonant decay
$\bs \to \phi K^+K^-$ and ``S--wave'' contamination \jpsifnot, is
neglected in the fit and its effect is accounted
for in the assigned systematic uncertainties. 
Using a large sample of Monte Carlo (MC) data 
the detector acceptance and the reconstruction requirements are
checked against biases
with a 0.2\% accuracy. 
The background asymmetries are consistent with zero, and the final
results for signal asymmetries are: $\au=(-0.7\pm6.4\stat\pm1.8\syst)\%$ and
$\av=(-12.0\pm6.4\stat\pm 1.6\syst)\%$. 
This measurement establishes a 
method to search for NP through \CP--violating observables in $P\to
VV$ decays without the
need of tagging and time--dependent analysis, which requires high
statistics samples. 

\section{Observation of the \jpsifnot\ decay}
The \jpsifnot\ decay has attracted significant attention as a 
potential ``S--wave'' contamination to the \jpsiphi\ signal when
the departure from the SM expectation of Tevatron measurement of mixed induced \CP-violation
was observed at level of about 1.5$\sigma$~\cite{sin2betas_PRL}. 
It was also suggested that enough signal of \jpsifnot\ decays
can be used to measure the \CP--violating phase \betas\ as well, without need of 
angular analysis~\cite{stone}. In addition, the \CP--odd nature of the \jpsi\fnot\ allows for
measuring the lifetime of the \bs\ \CP--odd eigenstate,
$1/\Gamma_s^{\rm{odd}}$, that is the lifetime of the heavy--mass eigenstate if \CP\ is conserved.
This year the Tevatron experiments have quickly confirmed the
observations of this mode~\cite{tev_fnot} from LHCb  and
Belle collaborations~\cite{lhcb_fnot}. In
the following we review the CDF result of the ratio of
branching fractions:
\begin{equation}
R=\frac{\Br(\jpsifnot)\Br(\fnotpipi)}{\Br(\jpsiphi)\Br(\phikk)},
\end{equation}
where $\Br(\fnotpipi)$ and $\Br(\phikk)$ are fixed to PDG values~\cite{PDG},
while $\Br(\jpsifnot)/\Br(\jpsiphi)$
is measured using 3.8 \ifb of data collected by the dimuon trigger. 

The sample selection is performed by a neural network trained to
maximize the separation between signal and background events. 
A threshold on the output of the neural network is chosen 
by maximizing $\epsilon/(2.5+\sqrt{N_b})$~\cite{punzi}, 
where $\epsilon$ is the signal reconstruction efficiency and $N_b$ 
is the number of background events estimated from mass distribution sidebands.
The background is dominated by a smooth combinatorial component. 
Physics backgrounds are studied using inclusive simulated decays of
 $b$--hadrons into \jpsi\ final states (fig.~\ref{fig:fnot}). 
The most prominent physics backgrounds in the
$\jpsi\pi\pi$ spectrum are $B^0\to \jpsi
K^{\star}(892)^0$ 
and $B^0\to \jpsi\pi^+\pi^-$ decays.

The ratio $\Br(\jpsifnot)/\Br(\jpsiphi)$ is evaluated as
$N(\jpsifnot)/N(\jpsiphi)\epsilon_{\rm{rel}}$, where $N(\jpsifnot)$ and
$N(\jpsiphi)$ are the number of signal events of \jpsifnot\ and
\jpsiphi\, respectively, extracted by performing an
unbinned extended maximum likelihood fit 
of the candidates mass distribution in [5.26;5.5] GeV/$c^2$ (fig.~\ref{fig:fnot}); 
$\epsilon_{\rm{rel}}$ is the relative efficiency for the reconstruction of the
two decays. The latter is evaluated by MC simulation, where \jpsiphi\
candidates are generated based on CDF preliminary results~\cite{new_sin2betas},\footnote{As a strong phase
  $\delta_\parallel$ is not measured we use the world average value
  from $B^0 \to \jpsi K^{0 \star}$ decays~\cite{PDG}.} while \jpsifnot\
candidates are modeled by a Flatt\'e distribution whose parameters
are fixed to the BES experiment results~\cite{BES}.
We found $N(\jpsifnot)=571\pm37\stat\pm25\syst$ with significance much greater than 5$\sigma$, 
and finally $R=0.292\pm0.020\stat\pm0.017\syst$. 
The measurement is in good agreement with 
determinations by other experiments and it is the most accurate result
to date. 
\begin{figure} 
\begin{center}
\begin{overpic}[width=0.38 \columnwidth]{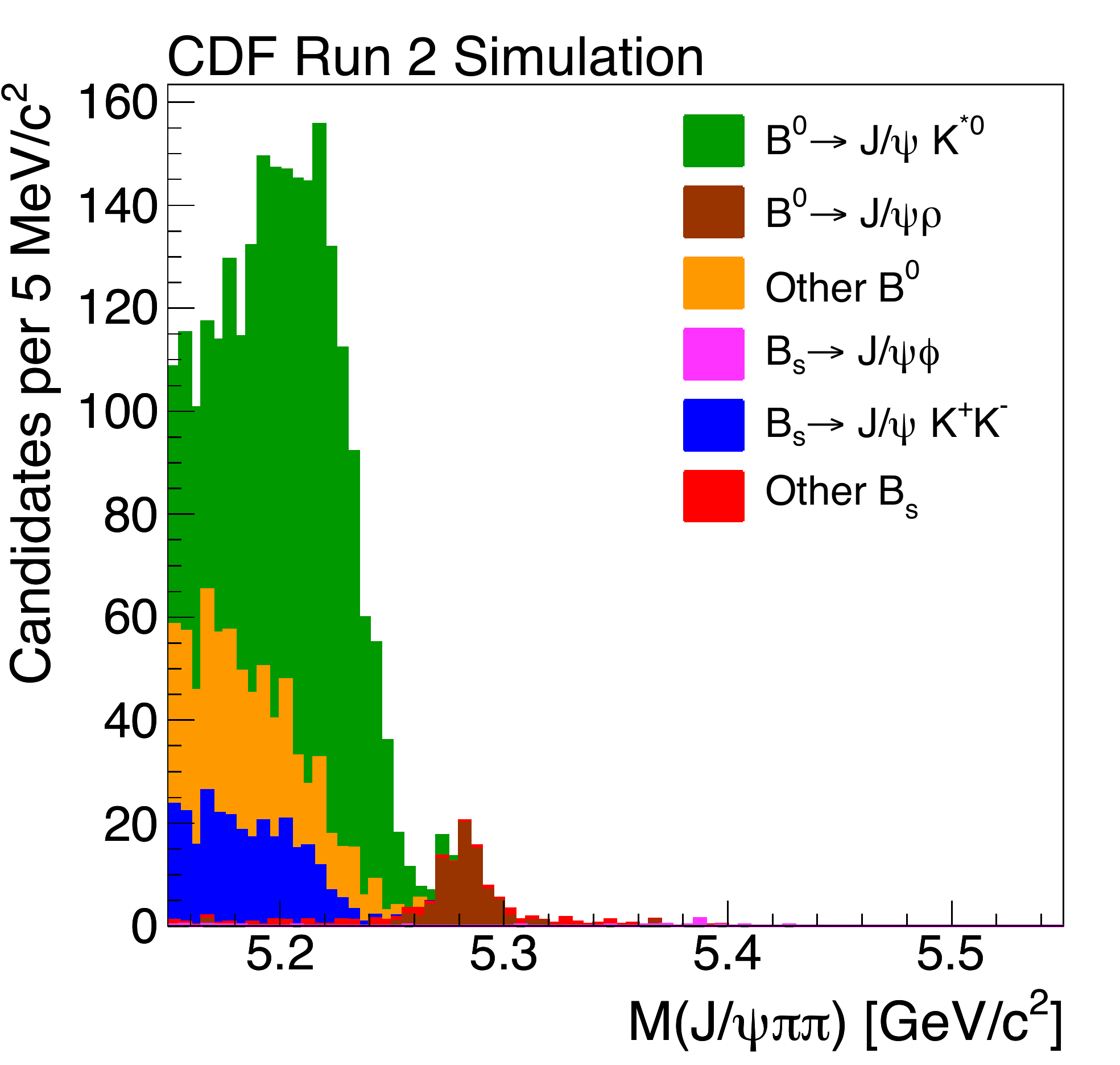}
\end{overpic}
\begin{overpic}[width=0.38 \columnwidth]{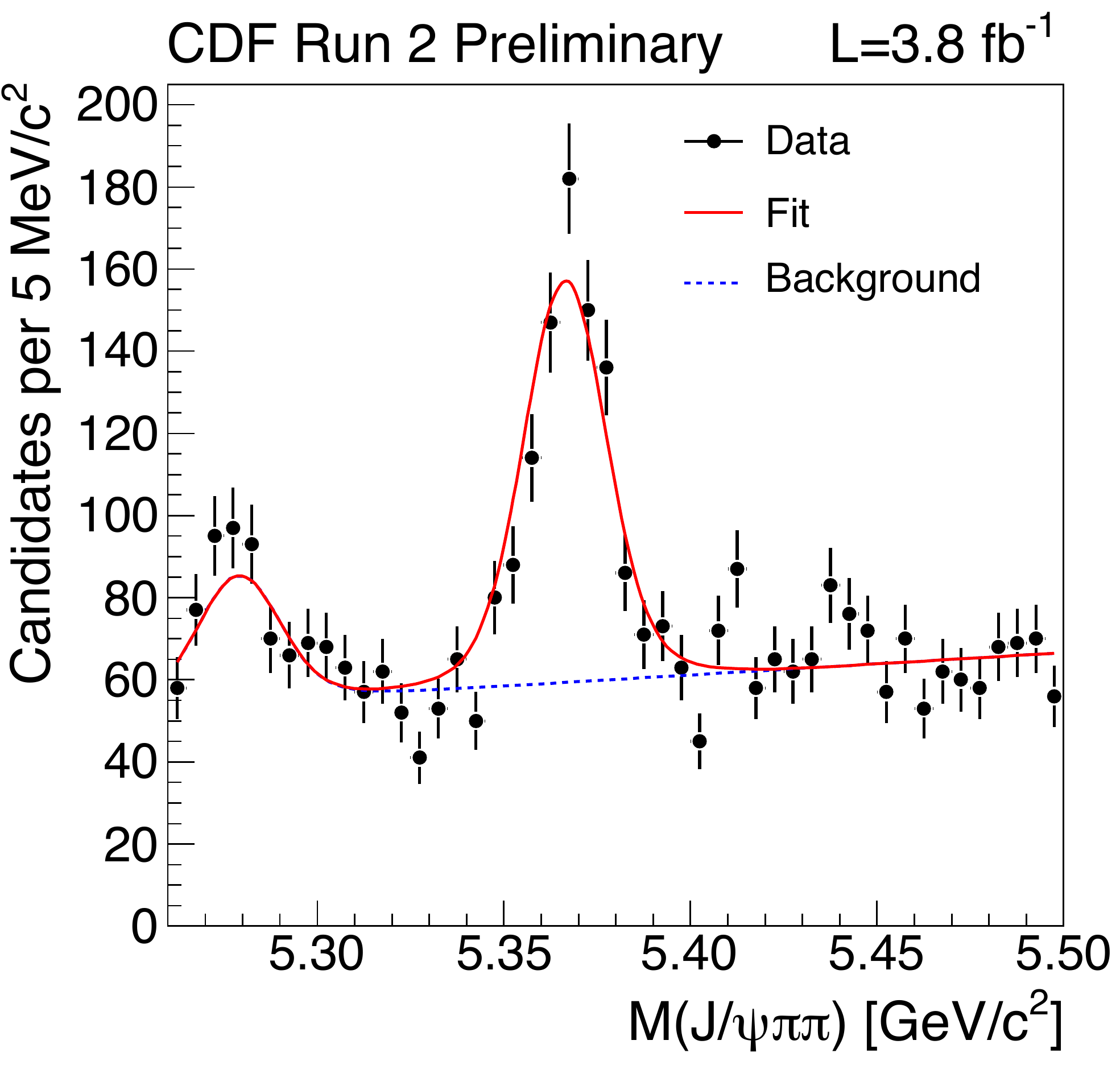}
\end{overpic}
\end{center}
\caption{\label{fig:fnot}
Left: stacked histogram of physics backgrounds in $\jpsi \pi\pi$ mass
distribution from simulation. Right: fit to the $\jpsi \pi\pi$ mass
distribution.}
\end{figure}

\section{First observation of \bsjpsik\ decays}
The two Cabibbo--suppressed decays
 \bsjpsikst\ and \bsjpsikS\ 
 allow disentanglement of penguin
contributions in the decays \jpsiphi\ and
$B^0\to\jpsi K^0_S$, respectively. 
The \bsjpsikst\ decay
could be used for the measurement of \DGs\ and polarization amplitudes,
and \bsjpsikS\ for measurement of $1/\Gamma_s^{\rm{odd}}$. 
The \bsjpsikS\ decay can also yield information on the $\gamma$ angle of the
unitarity triangle~\cite{gamma}. 

The CDF collaboration has recently reported the first observation of
these modes and the measurement of their branching ratios in 5.9 \ifb of data
selected by the dimuon trigger~\cite{newBsCDF}.
The sample selection was optimized maximizing the sensitivity 
for either finding evidence of a signal at 3$\sigma$, or excluding it at the same
confidence level; for the \bsjpsikS\ the selection is based on a
neural network, while for \bsjpsikst\ a simpler cut--based analysis is performed, in
both cases exploiting vertexing and kinematic discriminating
variables. Both analyses have two common background contributions:
 the combinatorial background and the partially reconstructed
background where  $\gamma$, $\pi$ or 
$K$ of multibody decays are not reconstructed. Other physics
backgrounds, such as $\Lambda_b
\to \jpsi \Lambda$ for \bsjpsikS\ or \jpsiphi\ for \bsjpsikst, give
negligible contributions.
A binned maximum likelihood fit to the mass distribution of the
candidates has been performed to extract the signal yields (fig.~\ref{fig:suppressed}): $64\pm14$
\bsjpsikS\  and $151\pm25$ \bsjpsikst\ signal events have been
observed, both with a significance greater than 5$\sigma$. 
Branching fractions are normalized to rates of the corresponding favored
modes, $\Br(B^0\to\jpsi K^0_S)$ and $\Br(B^0\to\jpsi K^{0 \star})$, and relative efficiency of reconstruction is evaluated by MC
simulation. The branching ratio of the favored decays are
fixed to their PDG values~\cite{PDG} and the fragmentation--fraction
$f_s/f_d$ is fixed to the most recent CDF
measurement of $f_s/(f_u+f_d)\Br(D_s \to \phi\pi)$~\cite{CDF_fsfd} combined with PDG
value of $\Br(D_s \to\phi\pi)$. Finally, the measured branching ratios are 
$\Br(\bsjpsikst)=(8.3 \pm 1.2\stat \pm 3.5\syst)\times 10^{-5}$ and 
$\Br(\bsjpsikS)=(3.5 \pm 0.6\stat \pm 0.6\syst)\times 10^{-5}$.
\begin{figure} 
\begin{center}
\begin{overpic}[width=0.45\columnwidth]{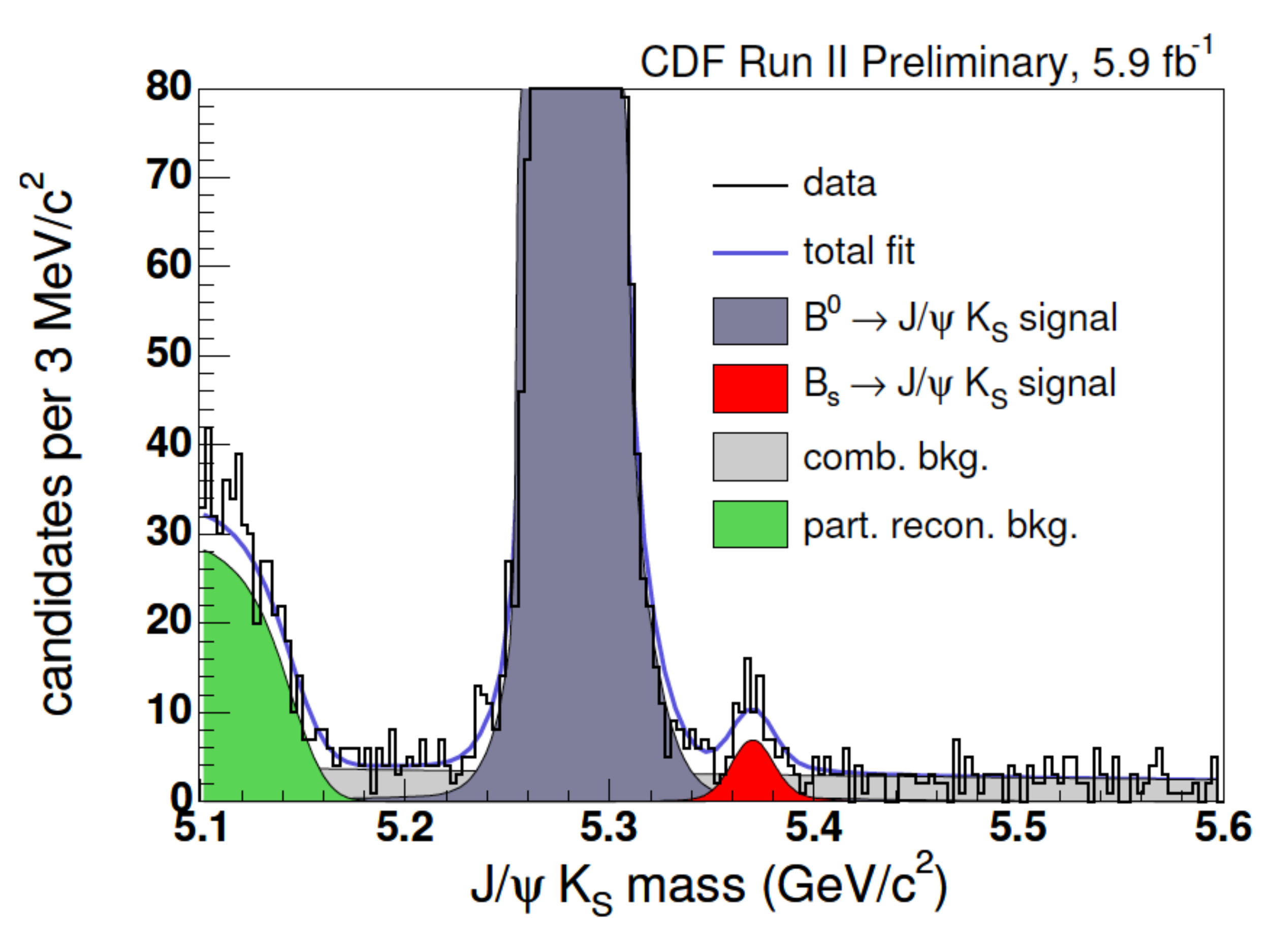}
\end{overpic}
\begin{overpic}[width=0.44\columnwidth]{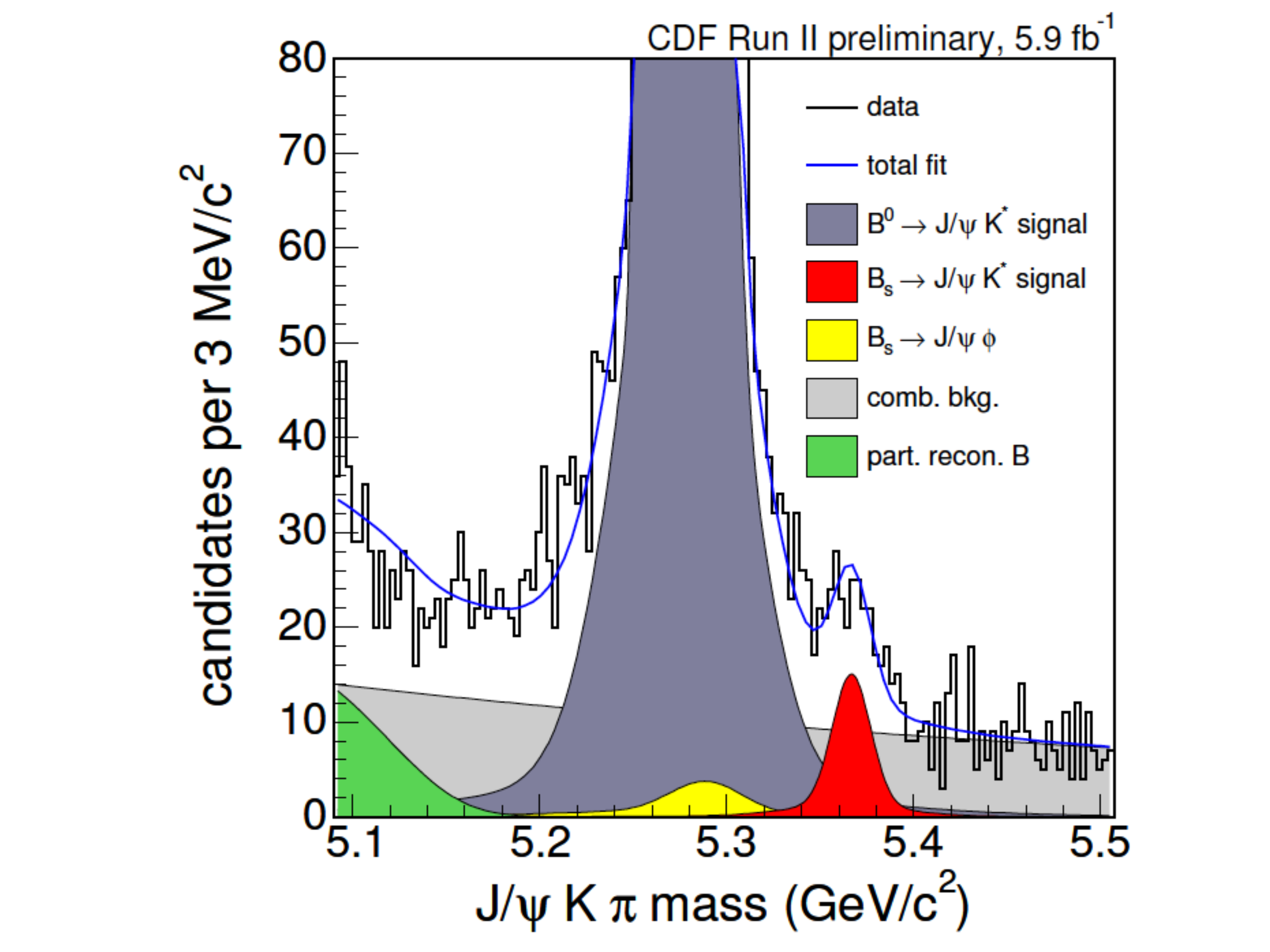}
\end{overpic}
\end{center}
\caption{\label{fig:suppressed}
Invariant mass distribution for $\jpsi K_S^0$ (left) and for $\jpsi
K\pi$ (right) with fit including the different contributions.}
\end{figure}


{\small
\begin{thebibliography}{99}
 \bibitem{CDFdetector} D. E. Acosta  {\it et al.}  (CDF collaboration),
   Phys. Rev. D{\bf 71}, 032001 (2005). 
 \bibitem{SVT} L. Ristori and G. Punzi,
  Annu. Rev. Nucl. Part. Sci. {\bf 60}, 595 (2010).
\bibitem{BVV_exp} P. Goldenzweig {\it et al.} (Belle collaboration), 
  Phys. Rev. Lett. {\bf 101}, 231801 
  (2008);\\  B. Aubert {\it et al.} (BaBar collaboration), 
  {\it ibid.} {\bf 99}, 201802 (2007).
\bibitem{BVV_th} M. Beneke {\it et al.}, Nucl. Phys. B{\bf 774}, 64 
  (2007); A. Datta {\it et al.}, Eur. Phys. J. C{\bf 60}, 279 
  (2009); A. Ali et al., Phys. Rev. D{\bf 76}, 074018 (2007); 
  X. Li {\it et al.}, Phys. Rev. D{\bf 71}, 019902 (2005).
\bibitem{sin2betas_PRL} T. Aaltonen {\it et al.} (CDF collaboration),
  Phys. Rev. Lett. {\bf 100}, 161802 (2008); 
 V. M. Abazov {\it et al.} (D0 collaboration), Phys. Rev. Lett. {\bf 101},
 241801 (2008).
\bibitem{phiphi_PRL} D. Acosta {\it et al.} (CDF collaboration),
  Phys. Rev. Lett. {\bf 95}, 031801 (2005).
\bibitem{phiphi} CDF collaboration, CDF Public Note 10064 and 10120
  (2010).
\bibitem{TP_th} A. Datta and D. London, Int. J. Mod. Phys. A{\bf 19}, 2505 (2004); A. Datta {\it et al.}, arXiv:1103.2442.
\bibitem{TP_CDF} CDF collaboration, CDF Public Note 10424 (2011).
\bibitem{PDG} K. Nakamura  {\it et al.} (Particle Data Group),
  J. Phys. G{\bf 37}, 075021 (2010).
\bibitem{stone} S. Stone and L. Zhang, arXiv:0909.5442.
\bibitem{tev_fnot} CDF collaboration, CDF Public Note 10404; I. Ripp-Baudot
  (for the D0 collaboration), PoS BEAUTY2011 (2011).
\bibitem{lhcb_fnot} R. Aaij {\it et al.} (LHCb collaboration),
  Phys. Lett. B{\bf 698}, 115 (2011); J. Li {\it et al.} (Belle collaboration),
  Phys. Rev. Lett. {\bf 106}, 121802 (2011).
\bibitem{punzi} G. Punzi, PHYSTAT-2003, MODT002 (2003), arXiv:physics/0308063.
\bibitem{new_sin2betas} CDF collaboration, CDF Public Note 10206.
\bibitem{BES} M. Ablikim {\it et al.} (BES collaboration),
  Phys. Lett. B{\bf 607}, 243 (2005).
\bibitem{gamma} R. Fleischer, Eur. Phys. J. C{\bf 10}, 299-306,
  (1999).
\bibitem{newBsCDF} T. Aaltonen {\it et al.} (CDF collaboration),
  Phys. Rev. D{\bf 83}, 052012 (2011). 
\bibitem{CDF_fsfd} T. Aaltonen {\it et al.} (CDF collaboration), Phys. Rev. D{\bf 77}, 072003 (2008). 
\end{thebibliography}
}
\end{document}